\documentclass[dvips]{article}
\usepackage{lscape,epsfig}
\usepackage{graphicx}
\usepackage{amssymb}
\usepackage{amsmath}
\textheight=22cm \DeclareSymbolFont{ppa}{OT1}{ppl}{m}{it}
\DeclareMathSymbol{\vv}{\mathalpha}{ppa}{'166}

minus 2.3mu \thickmuskip = 2.6mu plus 2mu minus 2.6mu

\begin{document}

\newcommand{\dd}{\,{\rm d}}
\newcommand{\ie}{{\it i.e.},\,}
\newcommand{\etal}{{\it et al.\ }}
\newcommand{\eg}{{\it e.g.},\,}
\newcommand{\cf}{{\it cf.\ }}
\newcommand{\vs}{{\it vs.\ }}
\newcommand{\zdot}{\makebox[0pt][l]{.}}
\newcommand{\up}[1]{\ifmmode^{\rm #1}\else$^{\rm #1}$\fi}
\newcommand{\dn}[1]{\ifmmode_{\rm #1}\else$_{\rm #1}$\fi}
\newcommand{\upd}{\up{d}}
\newcommand{\uph}{\up{h}}
\newcommand{\upm}{\up{m}}
\newcommand{\ups}{\up{s}}
\newcommand{\arcd}{\ifmmode^{\circ}\else$^{\circ}$\fi}
\newcommand{\arcm}{\ifmmode{'}\else$'$\fi}
\newcommand{\arcs}{\ifmmode{''}\else$''$\fi}
\newcommand{\MS}{{\rm M}\ifmmode_{\odot}\else$_{\odot}$\fi}
\newcommand{\RS}{{\rm R}\ifmmode_{\odot}\else$_{\odot}$\fi}
\newcommand{\LS}{{\rm L}\ifmmode_{\odot}\else$_{\odot}$\fi}

\newcommand{\Abstract}[2]{{\footnotesize\begin{center}ABSTRACT\end{center}
\vspace{1mm}\par#1\par \noindent {~}{\it #2}}}

\newcommand{\TabCap}[2]{\begin{center}\parbox[t]{#1}{\begin{center}
  \small {\spaceskip 2pt plus 1pt minus 1pt T a b l e}
  \refstepcounter{table}\thetable \\[2mm]
  \footnotesize #2 \end{center}}\end{center}}

\newcommand{\TableSep}[2]{\begin{table}[p]\vspace{#1}
\TabCap{#2}\end{table}}

\newcommand{\FigCap}[1]{\footnotesize\par\noindent Fig.\  %
  \refstepcounter{figure}\thefigure. #1\par}

\newcommand{\TableFont}{\footnotesize}
\newcommand{\TableFontIt}{\ttit}
\newcommand{\SetTableFont}[1]{\renewcommand{\TableFont}{#1}}

\newcommand{\MakeTable}[4]{\begin{table}[htb]\TabCap{#2}{#3}
  \begin{center} \TableFont \begin{tabular}{#1} #4
  \end{tabular}\end{center}\end{table}}

\newcommand{\MakeTableSep}[4]{\begin{table}[p]\TabCap{#2}{#3}
  \begin{center} \TableFont \begin{tabular}{#1} #4
  \end{tabular}\end{center}\end{table}}

\newenvironment{references}%
{ \footnotesize \frenchspacing
\renewcommand{\thesection}{}
\renewcommand{\in}{{\rm in }}
\renewcommand{\AA}{Astron.\ Astrophys.}
\newcommand{\AAS}{Astron.~Astrophys.~Suppl.~Ser.}
\newcommand{\ApJ}{Astrophys.\ J.}
\newcommand{\ApJS}{Astrophys.\ J.~Suppl.~Ser.}
\newcommand{\ApJL}{Astrophys.\ J.~Letters}
\newcommand{\AJ}{Astron.\ J.}
\newcommand{\IBVS}{IBVS}
\newcommand{\PASP}{P.A.S.P.}
\newcommand{\Acta}{Acta Astron.}
\newcommand{\MNRAS}{MNRAS}
\renewcommand{\and}{{\rm and }}
\section{{\rm REFERENCES}}
\sloppy \hyphenpenalty10000
\begin{list}{}{\leftmargin1cm\listparindent-1cm
\itemindent\listparindent\parsep0pt\itemsep0pt}}%
{\end{list}\vspace{2mm}}

\def\TYLDA{~}
\newlength{\DW}
\settowidth{\DW}{0}
\newcommand{\dw}{\hspace{\DW}}

\newcommand{\refitem}[5]{\item[]{#1} #2%
\def\REFARG{#3}\ifx\REFARG\TYLDA\else, {\it#3}\fi
\def\REFARG{#4}\ifx\REFARG\TYLDA\else, {\bf#4}\fi
\def\REFARG{#5}\ifx\REFARG\TYLDA\else, {#5}\fi.}

\newcommand{\Section}[1]{\section{#1}}
\newcommand{\Subsection}[1]{\subsection{#1}}
\newcommand{\Acknow}[1]{\par\vspace{5mm}{\bf Acknowledgements.} #1}
\pagestyle{myheadings}

\newfont{\bb}{ptmbi8t at 12pt}
\newcommand{\xrule}{\rule{0pt}{2.5ex}}
\newcommand{\xxrule}{\rule[-1.8ex]{0pt}{4.5ex}}
\def\thefootnote{\fnsymbol{footnote}}

\begin{center}
 {\Large\bf
  A New Lower Main Sequence Eclipsing Binary with Detached
  Components}\footnote{Based on photometric observations performed at
  CTIO, LCO and SAAO, and spectroscopic data collected with the Very
  Large Telescope at ESO Paranal Observatory under programme 382.D-0439(A).}
  \vskip1cm
  {\large
        M.~~R~o~z~y~c~z~k~a$^1$,
      ~~J.~~K~a~l~u~z~n~y$^1$,
      ~~P.~~P~i~e~t~r~u~k~o~w~i~c~z$^{1,2}$,
      ~~W.~~P~y~c~h$^1$,
      ~~B.~~M~a~z~u~r$^1$,
      ~~M.~~C~a~t~e~l~a~n$^2$
      ~~and~~I.~B.~~T~h~o~m~p~s~o~n$^3$\\
  }
  \vskip3mm {
  $^1$Nicolaus Copernicus Astronomical Center, ul. Bartycka 18, 00-716 Warsaw, Poland\\
     e-mail: (mnr, jka, pietruk, pych, batka)@camk.edu.pl\\
  $^2$Departamento de Astronom\'{\i}a y Astrof\'{\i}sica, Pontificia Universidad
      Cat\'olica de Chile, Av. Vicu{\~n}a Mackenna 4860, Macul, Santiago,
      Chile\\
     e-mail: (pietruk, mcatelan)@astro.puc.cl}\\
  $^3$The Observatories of the Carnegie Institution of Washington, \\813 Santa Barbara
      Street, Pasadena, CA91101, USA\\
     e-mail: ian@ociw.edu
\end{center}
\Abstract { We present an analysis of NGC2204-S892 -- a new detached
eclipsing binary composed of two late K dwarfs. Based on three
photometric campaigns launched in 2008 we obtained 5 light curves (3
in $V$, 1 in $B$ and 1 in $I$), and derived an orbital period of
0.451780$\pm$0.000001~d. We also obtained 20 VLT/UVES spectra,
enabling accurate radial velocity measurements. The derived masses
and radii of the components ($m_1 = 0.733 \pm 0.005$~$M_\odot$ and
$R_1 = 0.72 \pm 0.01$~$R_\odot$; $m_2 = 0.662 \pm 0.005$~$M_\odot$
and $R_2 = 0.68 \pm 0.02$~$R_\odot$) are consistent with the
empirical mass-radius relationship established recently for lower
main sequence stars in binary systems; in particular we find that
both stars are oversized compared to theoretical models.
NGC2204-S892 is very active: both components show variable emission
in H$\alpha$ and H$\beta$ and are heavily spotted, causing the light
curve to show appreciable changes on a timescale of weeks. Our
results add to the increasing evidence that the observed inflation
of the radii of K and M stars is related to high levels of magnetic
activity. }
{\bf Key words:} {\it binaries: eclipsing -- stars: individual:
NGC2204-S892 -- stars: low-mass -- stars: K-type}
\section {Introduction} \label{sect:intro}
Significant discrepancies are known to exist on the lower main
sequence (LMS) between theoretical predictions and the actual
parameters of stars. For stars in the 0.2-0.8 $M_\odot$ mass range
observed radii are larger by up to 10 -- 15 percent, and effective
temperatures lower by several percent than predicted by theory.
According to Morales, Ribas, \& Jordi (2008), ``... current stellar
structure and evolution models are not adequate to describe the
physical (radii) and radiative (effective temperature, color
indices) properties of... low-mass stars. The problem is
particularly severe for the analysis of open clusters or star
forming regions''. Indeed, LMS stars (LMSS) appear 50 -- 90 percent
older or younger in mass-radius diagrams than they really are,
depending on whether post-- or pre--MS models are used (Stassun et
al. 2009). No definitive explanation for the inflated radii of LMSS
has yet been given, but mounting evidence suggests that they result
from intense magnetic activity associated with high spot-coverage
(Chabrier \etal 2007; Ribas et al. 2008; Morales et al. 2008). A
positive correlation between radius and activity level was found for
a few members of close binaries by L\'opez-Morales (2007). This
important discovery clearly deserves further study, extending it
onto a larger region of parameter space.

Active low-mass stars are fast rotators with rotation periods
shorter than $\sim3$ days, and their activity is strongly correlated
with rotational velocity ({\eg } Mohanty \& Basri 2003). It is
evident, then, that the best objects with which to study the
dependence of radius on activity are members of close eclipsing
binary systems. First, their masses and radii can be accurately
determined from photometric and spectroscopic data. Second, they are
rotationally synchronized, and due to the spread in orbital periods
they should exhibit a wide range of activity. Unfortunately, LMS
binaries are very rare: according to Shaw \& L\'opez-Morales (2007),
there is only one such system per about million stars listed in
recent sky surveys; whereas Ribas et al. (2008) mention only 7
binaries for which masses and radii have been derived with
uncertainties below 3 percent. The situation may improve owing to
future large-area synoptic surveys like SDSS II, Pan-STARRS and LSST
(Blake et al. 2008), but at the moment every new LMS binary with
components in the mass range 0.2 $M_\odot < m <$ 0.8 $M_\odot$ is
highly valuable as a potential source of data enabling detailed
studies of the effects of magnetic activity on stellar structure and
evolution.

It should be mentioned here that optical interferometry has recently
enabled accurate direct measurements of the radii of nearby LMSS.
However, the masses are not directly measurable in this case, and
can only be estimated through comparison with empirical
mass-luminosity relations or other indirect methods. Moreover,
interferometric measurements are limited to single stars whose
rotational velocities rarely exceed 3 km s$^{-1}$, implying a low
activity or lack thereof. Not surprisingly, the radii of single M
dwarfs measured by interferometry do not show obvious excesses, and
within observational errors they are compatible with models (Demory
et al. 2009). Those authors detected possibly meaningful deviations
only for early K stars, and found them to disappear when the mixing
length parameter $l_{\rm mix}/H_P$ of 1.5--1.9 is used instead of
1.0.

In the present paper we analyze NGC2204-S892 -- a newly discovered detached
double-lined eclipsing binary composed of two late K dwarfs
(hereafter: S892). The binary is a foreground object in the field of
the open cluster NGC~2204, originally described by Rozyczka et al.
(2007) as star \#892. They reported orbital period of slightly
less than 0.5 d with brightness varying between $V\approx17.4$~mag
at maximum and $V\approx18.4$~mag at primary minimum, and secondary
minimum $\sim$0.55 mag deep. Based on the dereddened $B-V$ color and the
difference between the depths of the minima, they estimated $\sim$0.6 $M_\odot$
for the mass of the primary.

In 2008 we collected 5 light curves of S892 (3 in $V$, 1 in $B$ and
1 in $I$) and derived an orbital period of 0.451780$\pm0.000001$ d.
We also obtained 20~UVES spectra enabling accurate measurements of
the radial velocity amplitudes (145$\pm$2~km~s$^{-1}$ and
165$\pm$2~km~s$^{-1}$ for primary and secondary, respectively). The
details of our observing runs are provided in Section~2. Section~3
presents the analysis of the collected data and the resulting
physical parameters of the binary. The properties of the binary are
discussed in Section~4 in relation to the discrepancies between
empirical and theoretical mass-radius relations for LMSS, and a
brief summary of our work is presented in Section~5.
\section {Observations} \label{sect:obs}
S892 is located in Canis Major at $\alpha_{2000}$ = $06^{\rm
h}15^{\rm m}$55\zdot$^{\rm s}$42 and $\delta_{2000}$ =
$-18^{\circ}44'51\zdot\arcs7$. We observed it for a week in January
2008 with the 2.5-m Ir\'en\'ee du Pont telescope at Las Campanas
Observatory (LCO), for another week in November 2008 with the 1.0-m
Yale telescope at Cerro Tololo Interamerican Observatory (CTIO), and
for 2 weeks in December 2008 with the 1.0-m Elisabeth telescope at
the South African Astronomical Observatory (SAAO). The observations
were made in Johnson $B$ and $V$, and Cousins $I$ bands, yielding a
total of 1282 frames. A journal of the observations is given in
Table \ref{tab:calendar} together with numbers of measurements
obtained in each run and filter. The spectra of the system were
taken between early October 2008 and mid-January 2009 with the UVES
spectrograph at VLT UT2 (Kueyen) at the ESO Paranal Observatory
under programme 382.D-0439(A).
\begin{table}[h!]
 \begin{center}
  \caption{\small  Journal of photometric observations \label{tab:calendar}}
  {\small
   \begin{tabular}{lccr}
    \hline
    Site& Observing period& Filters& No. of frames\\
    \hline
    LCO (2.5m)& 11.01-17.01.2008& $B/V$& 85/237\\
    CTIO (1.0m) & 05.11-13.11.2008& $V$& 464\\
    SAAO (1.0m) & 03.12-16.12.2008& $V/I$& 331/193\\
    \hline
   \end{tabular}
  }
 \end{center}
\end{table}
\subsection {Photometry}\label{sect:phot}
\subsubsection{LCO}
Images were taken in the $B$ and $V$ bands with the TEK5 camera on
the 2.5-m Ir\'en\'ee du~Pont telescope at a scale of 0.259
arcsec/pixel. The FoV was 8.8$\times$8.8 arcmin$^2$, and exposure
times ranged from 50 to 120~s in $V$ and from 120 to 180~s in $B$.
In a total of $\sim$17.5 hours of monitoring time 85 frames in $B$
and 237 in $V$ were collected. All $B$ frames and 235 of the $V$
frames were useful. The PSF had a FWHM ranging from 0.7 to 2.3
arcsec in $B$ and from 0.6 to 2.2 arcsec in $V$, with median values
of 1.1 and 1.0 arcsec, respectively.
\subsubsection{CTIO}
Images were taken with the Y4K camera on the Yale 1.0-m telescope in
$V$ band only, at a scale of 0.289 arcsec/pixel. The FoV was
19.7$\times$19.7 arcmin$^2$, and every frame was exposed for 100~s. In a
total of $\sim$41 hours of monitoring time 464 frames were
collected, 459 of which were useful. Only the FoV quadrant
containing the binary was processed through the reduction pipeline.
In this quadrant the PSF had a FWHM ranging from 1.1 to 2.2 arcsec, with a
median of 1.6 arcsec.
\subsubsection{SAAO}
Images were taken in the $V$ and $I$ bands with the STE4 camera on
the 1.0-m Elisabeth telescope at a scale of 0.31 arcsec/pixel.
The FoV was 5.3$\times$5.3 arcmin$^2$, and exposure times ranged
from 100 to 200~s in $V$ and from 70 to 150~s in $I$. In a total of
$\sim$37 hours of monitoring time 331 frames in $V$ and 193 in $I$
were collected, out of which 322 and 181 were useful, respectively.
The seeing was generally poor, and the resulting PSF had a FWHM
ranging from 1.3 to 3.6 arcsec in $V$ and from 1.2 to 3.3 arcsec in
$I$, with median values of 2.3 and 1.9 arcsec, respectively.
\subsubsection{Data reduction}\label{sect:photred}
The preliminary processing of the raw data was performed under
IRAF\footnote {IRAF is distributed by the National Optical Astronomy
 Observatories, which are operated by the AURA, Inc., under cooperative
 agreement with the NSF.
}. All frames were de-biased and flat-fielded with median-averaged
sky flats. The photometry was performed with the
DAOPHOT/ALLSTAR package (Stetson 1987). The reduction
procedure started from the identification of stars with the
subroutine FIND, followed by aperture photometry with the
subroutine PHOT. A PSF varying
quadratically with $(x,y)$ coordinates was then constructed
for each frame based on 30-40 isolated stars, and used for
profile photometry with the ALLSTAR
subroutine. The images were inspected visually, and for each camera
and each filter one of the best frames was chosen as a template.
Instrumental magnitudes of template stars were transformed to the
standard system using the data of Kassis \etal (1997) available in
the WEBDA database. For each frame taken with the same camera and
filter the template stars were identified by means of
transformations of $(x,y)$ coordinates, and ALLSTAR
photometry for those frames was transformed to the standard system
with the template serving as an intermediary. The quality of the
transformation was verified by plotting the light curve of a
comparison star with similar colors which was located close to
S892 in the frames: it did not show any secular trend or
discontinuity. The mean $V$-band magnitude of that comparison star calculated
from all 2008 observations was $V_{\rm comp}=17.627$ mag. The rms
deviation from $V_{\rm comp}$ was 0.006, 0.017 and 0.016 mag for
LCO, CTIO and SAAO data, respectively, and in all cases it was
practically equal to the mean error of the photometry calculated by
DAOPHOT. The corresponding mean magnitudes obtained from the three
sets of data differed from $V_{\rm comp}$ by less than $10^{-3}$
mag.

\begin{figure}[h!]
 \begin{center}
  \includegraphics [width=\textwidth, bb = 54 358 565 690, clip] {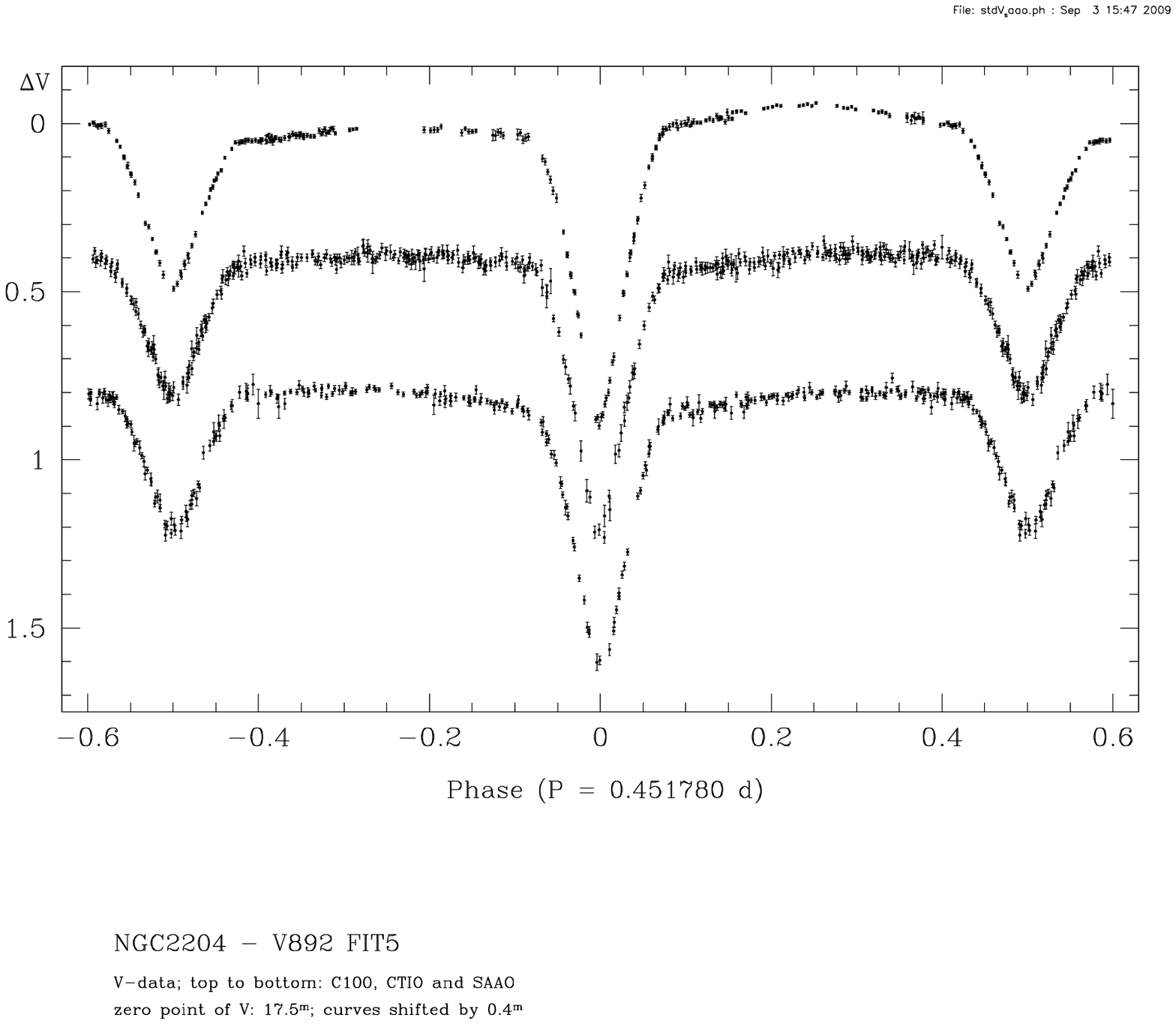}
  \caption {Phased $V$-band light curves obtained at (from top to bottom):
   LCO, CTIO and SAAO. The errors, with mean values of 0.007, 0.017 and 0.014~mag,
   respectively, are calculated by DAOPHOT. The zero-point on the vertical
   axis corresponds to 17.5 mag, and the curves are displaced from one another by 0.4~mag.
}
  \label{fig:Vdata}
 \end{center}
 \vskip -5mm
\end{figure}

Based on all 2008 data in the $V$-band we derived the following
ephemeris of the system:
\begin{eqnarray}
 t_0 &=& {\rm HJD}~2454480.6326\pm~{\rm 0.0001}
 \label{eq:ephem}\\
 P   &=& 0.451780\pm0.000001~{\rm d} \nonumber
\end{eqnarray}
The resulting phased $V$-band light curves are shown in Fig. \ref{fig:Vdata}.
Their variability can be immediately seen by comparing the LCO curve with
any of the remaining two. Even the CTIO and SAAO curves differ noticeably,
although they are separated by just one month in time.
\subsection {Spectroscopy} \label{sect:spectr}
UVES spectra of S892 were obtained in service mode on 10 nights
spaced irregularly between 03.10.2008 and 18.01.2009. We used the
red arm of the spectrograph in a standard setting centered at 5800\AA
with the detectors binned 2$\times2$. S892 is rather faint, but
its short orbital period sets stringent limits on the exposure time
(long exposures would blur all spectral features). To keep both S/N
and effective resolution as high as possible we chose a $t_{\rm exp}$ of
1200 s (\ie 3\% of the orbital period). In practice two spectra
exposed for 600 s each were taken one after another, and
combined during data reduction. Such a procedure
yielded an average $S/N\approx18$ in the reduced spectra, and also
facilitated the removal of cosmic rays.

On every observing night four 600 s exposures of S892 were taken in
the following sequence: exposure, ThAr calibration spectrum, exposure,
flat-field image, exposure, ThAr calibration spectrum, exposure,
flat-field image. Each spectrum was delivered in two segments recorded
by independent CCD's: the ``blue" one extending from 4780 to 5755{\AA}
and the ``red" one extending from 5835 to 6800{\AA}.
We obtained 20 blue and 20 red spectra altogether, at orbital
phases between 0.18 and 0.97, with most of them grouped around
quadratures. The reduction of the data was performed under IRAF. The
raw frames were debiased and flatfielded, and the spectra were
extracted and callibrated with the help of NOAO, IMRED and ECHELLE
packages.

Radial velocity measurements were based on the broadening function
(BF) formalism discussed by Rucinski (2002). The main cause of line
broadening in S892 is fast rotation of the components,
contributions from other factors are practically negligible. For a
rigidly rotating spherical star we have
\begin{eqnarray}
 {\rm BF}_{rot}(v) &=&A \left[(1-\beta)\sqrt{1-a^2} + {\frac{\pi}{4}}
  \beta (1 - a^2)\right]
  \label{eq:bfrot}\\
 {\rm with}~~~~a&=&{\frac{v-v_{rad}}{v_{rot}\sin{i}}}~, \nonumber
\end{eqnarray}
where BF$_{rot}$ is the rotational BF, $v_{rad}$ and $v_{rot}$ are,
respectively, radial and rotational velocity, $\beta$ is the
coefficient of linear limb darkening, and $A$ is an amplitude. The
observed BFs were extracted separately from blue and red segments of
all spectra, using the spectrum of K5V star HD 10361 from the UVES
library as a template. Model BFs used for our analysis were
convolutions of two rotational BFs given by equation
(\ref{eq:bfrot}) and a Gaussian with a standard deviation of
40~km~s$^{-1}$.  Fits to the observed BFs were performed by means of
nonlinear least-squares procedures from the GNU Scientific Library.
The resulting radial velocities listed in Table \ref{tab:radvel} are
averages from $v_{\rm blue}$ and $v_{\rm red}$, \ie fits to blue and
red segment of the spectrum (the errors are estimated from the
formula $\sigma_v = |v_{\rm blue}-v_{\rm red}|/\sqrt{2}$). As a
by-product we obtained projected rotational velocities of both
components of the S892 system: $v_{\rm 1}\sin{i}= 79.2 \pm
5.9$~km~s$^{-1}$ for the primary and $v_{\rm 2}\sin{i}= 76.2 \pm
8.8$~km~s$^{-1}$ for the secondary. We note that the accuracy of the
velocity measurements is degraded by noticeable distortions of the
observed BFs due to stellar spots. Not surprisingly, the
observational errors are significantly and systematically larger for
the secondary component whose lines are weaker and more difficult to
measure.
\begin{table}
 \begin{center}
  \caption{\small  Radial velocities.\label{tab:radvel}}
  {\small
   \begin{tabular}{ccrr}
    \hline
    HJD-2454000 & Phase  & $v_{\rm 1}$ [km s$^{-1}$]& $v_{\rm 2}$ [km s$^{-1}$]\\
    \hline
    742.8072& 0.3146& $-131.6 \pm 2.0$&  $149.4 \pm~0.9$\\
    742.8239& 0.3518& $-117.1 \pm 4.2$&  $132.7 \pm~6.7$\\
    744.7988& 0.7231&  $145.6 \pm 0.4$& $-155.5 \pm~5.5$\\
    744.8155& 0.7600&  $149.0 \pm 2.6$& $-157.7 \pm~4.4$\\
    746.8112& 0.1774& $-128.4 \pm 5.7$&  $145.7 \pm~6.3$\\
    746.8279& 0.2144& $-141.5 \pm 2.1$&  $159.5 \pm~0.4$\\
    773.7041& 0.7041&  $140.4 \pm 1.4$& $-151.9 \pm10.9$\\
    773.7167& 0.7319&  $148.5 \pm 3.6$& $-156.1 \pm~9.9$\\
    775.7511& 0.2351& $-141.8 \pm 1.4$&  $158.6 \pm~5.2$\\
    775.7679& 0.2722& $-142.6 \pm 3.5$&  $159.2 \pm~5.0$\\
    792.7760& 0.9191&   $73.6 \pm 0.2$&  $-73.0 \pm~1.8$\\
    792.7928& 0.9562&   $48.3 \pm 2.8$&  $-54.8 \pm~5.2$\\
    793.8002& 0.1860& $-132.9 \pm 4.4$&  $151.6 \pm~4.5$\\
    793.8169& 0.2230& $-142.2 \pm 3.9$&  $160.9 \pm~5.6$\\
    796.7807& 0.7832&  $146.6 \pm 6.8$& $-161.8 \pm~5.3$\\
    796.7974& 0.8202&  $138.9 \pm 0.5$& $-145.5 \pm~4.9$\\
    798.6566& 0.9354&   $60.4 \pm 1.6$&  $-66.3 \pm~1.7$\\
    798.6737& 0.9734&   $40.6 \pm 0.9$&  $-30.7 \pm~7.8$\\
    849.5991& 0.6950&  $140.6 \pm 1.4$& $-151.2 \pm~8.9$\\
    849.6159& 0.7323&  $148.9 \pm 0.1$& $-160.0 \pm12.6$\\
    \hline
   \end{tabular}
  }
 \end{center}
 \vskip -2mm
 {\footnotesize
 Barycentric radial velocities of S892 measured from the UVES data
 and phased according to the ephemeris given by equation (\ref{eq:ephem}).
 }
 \vskip -3mm
\end{table}
\section {Modeling} \label{sect:mod}
We modeled the S892 system using the PHOEBE interface (Pr\v sa \&
Zwitter 2005) to the Wilson-Devinney code (Wilson \& Devinney 1971).
Because the narrow eclipses in Fig. \ref{fig:Vdata} indicate that
the components of the binary are far from contact, a detached
configuration was adopted for all calculations. However, since the
orbital period of S892 is rather short, and its light curves suggest
that both components are nonspherical, we also included proximity
effects for the primary and secondary. The bolometric albedo was
kept fixed at 0.5, which is the standard value for stars with
convective envelopes, and we enabled the double reflection option.
For the limb darkening a logarithmic law was used as implemented in
PHOEBE 0.31a. In all calculations the gravity brightening
coefficient $\beta_1$ was set to 0.32 -- the classical value
obtained by Lucy (1967). To check the sensitivity of the results to
that parameter, a few converged fits were re-run with $\beta_1=0.2$
-- a value obtained by Claret (2000). Changes, if any, were of the
same order as the accuracy of the fit.

As there was no indication for nonzero eccentricity in any of our
five light curves, we set $e\equiv0$. The period was adopted from
the photometric data. Both components were assumed to be entirely
synchronized, \ie their synchronicity parameters were set to 1.0.
Having no clues about the chemical composition of S892 we assumed
that it has solar metallicity. Rozyczka \etal (2007) estimated that
the system is located not more than $\sim$1 kpc away from the Sun
and $\sim$250~pc below the Galactic plane, so that it most probably
belongs to the thin disk population, and its true metallicity should
not differ much from the assumed value. We re-ran a few converged
fits with [Fe/H]~=~-0.5 and [Fe/H]~=~0.5. No significant changes
were found compared to the case with [Fe/H]~=~0.

The estimates of Rozyczka \etal (2007) were used as the initial
guess for the masses of the components. Based on masses and period,
we obtained an estimate of the semimajor axis $a$ of the system. An
initial guess for the inclination $i$ was provided by a preliminary
fit to the SAAO $V$-band curve, which was the most symmetric one
among the three. A preliminary fit to the velocity curve in turn
yielded updated (\ie. compatible with $i$) approximations for masses
and $a$. We checked that keeping $T_{\rm 1}$ at a value fixed
between 4100 K and 4300 K does not significantly influence the
outcome of iterations for any parameter of the system except $T_{\rm
2}$, and based on our approximate value of $m_{\rm 1}= 0.75$
$M_\odot$ we adopted $T_{\rm 1}=4200$ K -- in agreement with $T_{\rm
2}=4220\pm150$ K obtained by Torres \& Ribas (2002) for the
secondary component of V818 Tau with $m_{\rm 2} = 0.76$ or $T_{\rm
2}=4200\pm200$ K obtained by Bayless \& Orosz (2006) for both
components of 2MASS J05162881+2607387 with $m_{\rm 1} = 0.79$ and
$m_{\rm 2}=0.77$ $M_\odot$.

\begin{figure}[b!]
 \begin{center}
  \includegraphics[width=\textwidth, bb = 40 163 565 690, clip] {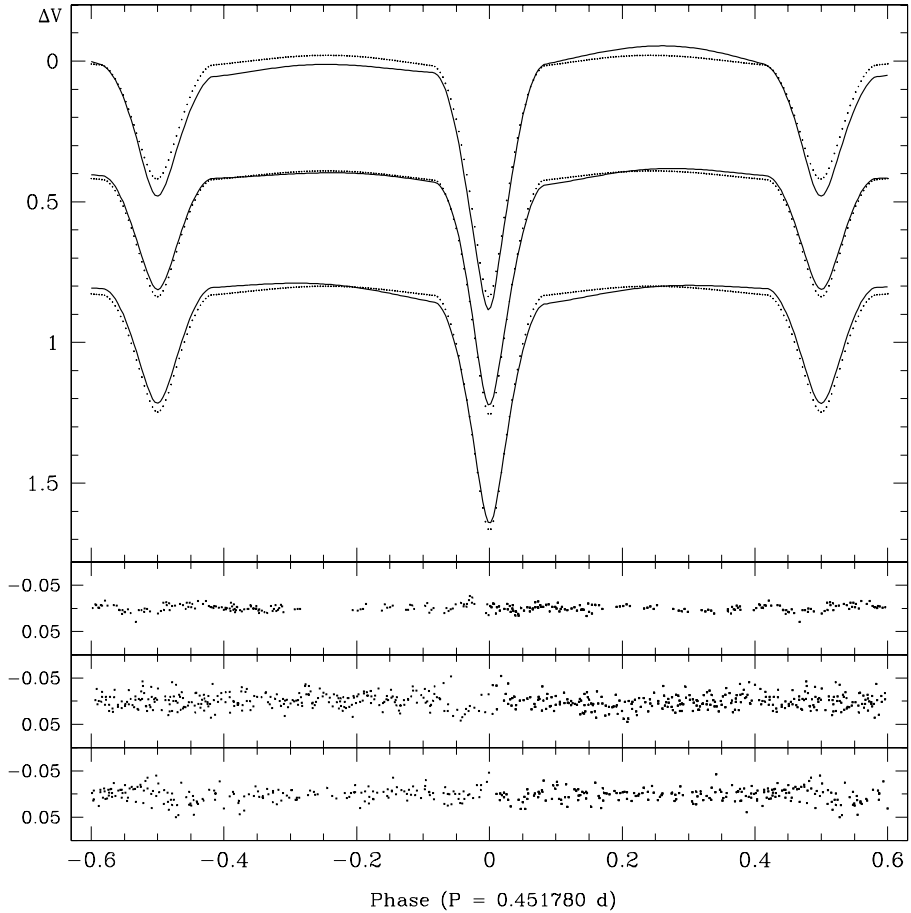}
  \vskip -2mm
  \caption {Upper panel: example fits to $V$-band light curves phased with $P=0.451280$
   d (solid lines). Dotted lines: the same fits shown after the spots have been
   removed from both components of the system. From top to bottom: LCO, CTIO and SAAO.
   The zero-point on the vertical axis corresponds to $V = 17.5$~mag, and the curves are
   displaced from one another by 0.4~mag. Lower panels: residuals from fits with spots.
   From top to bottom: LCO, CTIO and SAAO. The base fit (see text for explanations)
   was obtained from LCO data.}
  \label{fig:Vfit}
  \vskip -5 mm
 \end{center}
\end{figure}

Because of the evident variability of the light curve (see Fig.
\ref{fig:Vdata}) we could not use the photometric data {\rm en
bloc}. Instead, we adopted the following procedure:
\begin{enumerate}
 \item
 Choose one of LCO, CTIO or SAAO photometric data sets.
 \item
 \vspace {-1.5mm}
 Keeping $a$ and mass ratio $q \equiv m_{\rm 2}/m_{\rm 1}$ fixed, make
 a fit to the light curve by iterating for $T_{\rm 2}$, $i$,
 primary's luminosity $L_{\rm 1}$ and surface potentials of both
 componets. Add spots to account for the asymmetries of light curves. The result
 will serve as the {\em base fit} for the next steps of the procedure.
 \item
 \vspace {-1.5mm}
 Fix all parameters of the base fit except $a$ and $q$. Make a fit to the
 velocity curve by iterating for $a$ and $q$.
 \item
 \vspace {-1.5mm}
 Go to (2). Repeat steps (2) -- (4) until the base fit converges
 upon completing (2). We consider the fit converged when stellar radii $R_{\rm 1}$
 and $R_{\rm 2}$  change by less than 0.01 $R_\odot$ during one iteration step.
 \item
 \vspace {-1.5mm}
 Fix all parameters of the converged base fit. Substitute the first photometric
 data set with one of the remaining two. Adjust number, location and parameters
 of spots until photometric residua do not show any systematic or large-scale
 variability in phase.
 \item
 \vspace {-1.5mm}
 Relax $a$ and $q$. Make a fit to the velocity curve.
 \item
 \vspace {-1.5mm}
 Fix $a$ and $q$. Relax the remaining parameters, and make a fit to the
 photometric data.
 \item
 \vspace {-1.5mm}
 Go to (6). Repeat steps (6) -- (8) until upon completing (7) the fit converges
 according to the same criterion as in (4).
 \item
 \vspace {-1.5mm}
 Repeat steps (5) -- (8) for the third set of photometric data.
 \item
 \vspace {-1.5mm}
 Repeat steps (2) -- (8) using another set of photometric data to obtain the
 base fit.
 \item
 \vspace {-1.5mm}
 Repeat steps (2) -- (8) using the last set of photometric data to obtain the
 base fit.
\end{enumerate}
An example result of fitting is shown in Fig. \ref{fig:Vfit}, with
observational points omitted for clarity. In this case the base fit
was obtained from the LCO data. The rms residual $V$ magnitude of
LCO, CTIO and SAAO data is equal to 0.007, 0.017 and 0.016 mag,
respectively; \ie it is practically the same as the accuracy of
photometric measurements reported in Section \ref{sect:photred}. The
rms residual $B$ and $I$ magnitudes amount to 0.014 and 0.017 mag,
respectively. Note that all $V$-band light curves in Fig. \ref{fig:Vfit}
are asymmetric (the same holds for the $B$- and $I$-band light
curves). The proximity effect is clearly visible in the
``de-spotted" light curves (dotted lines in Fig. \ref{fig:Vfit}),
indicating a significant tidal distortion of the components of S892.

The number of spots placed on model stars to match the photometric
data varied from 1 to 5 per star in different fits. Both dark and
bright spots were introduced, with $|\Delta T_{\rm eff}|/T_{\rm
eff}$ ranging from 0.01 to 0.06. In most cases the spot

\begin{figure}[t!]
 \begin{center}
  \includegraphics[width=\textwidth, bb = 36 358 565 690, clip] {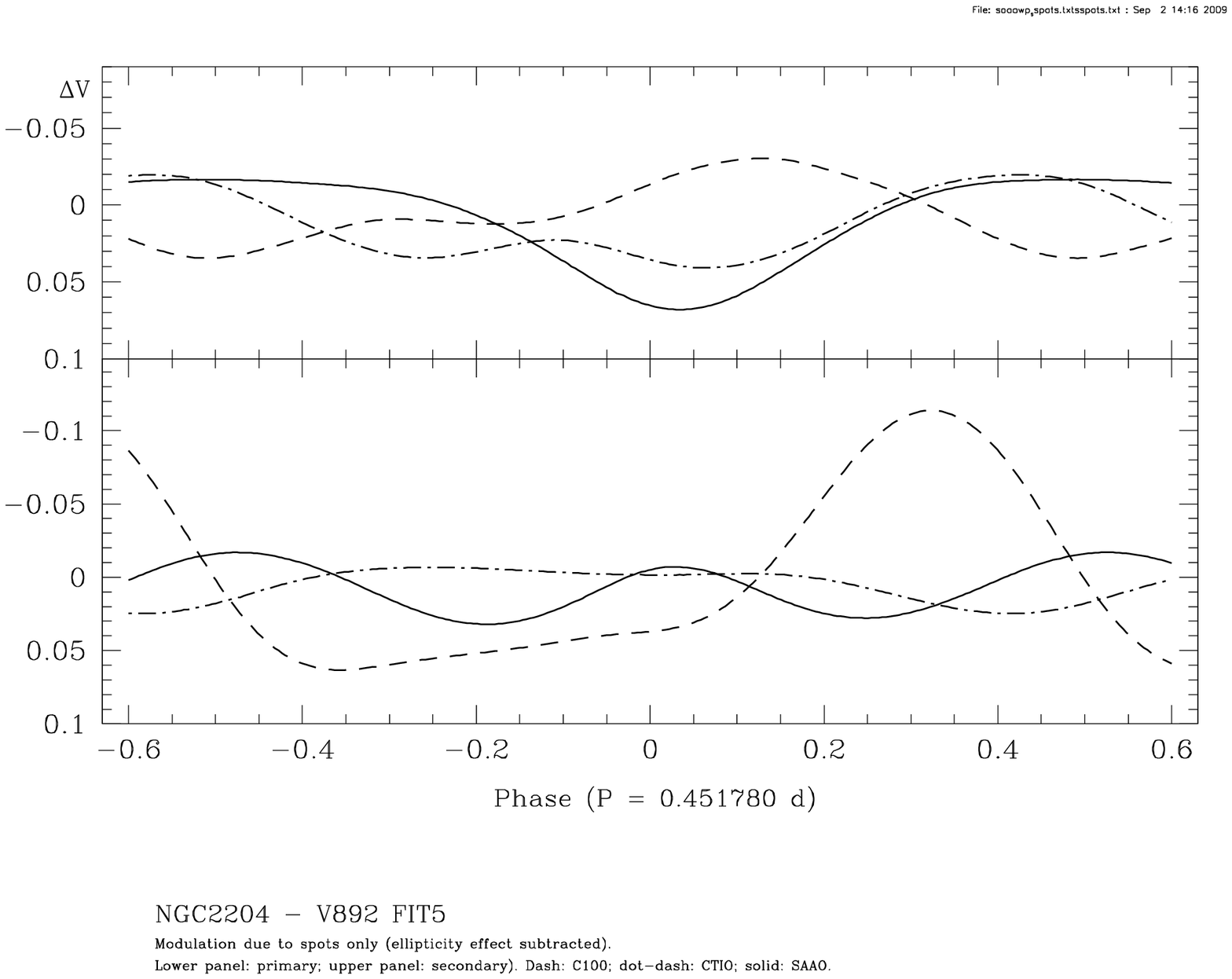}
  \caption {Example modulations of $V$-band magnitude due to spots on primary and
   secondary component of S892 (upper and lower panel, respectively). Dashed,
   dash-dotted and solid lines: fits to LCO, CTIO and SAAO light curves. }
  \label{fig:modul}
 \end{center}
 \vskip -5mm
\end{figure}

\noindent coverage was rather high -- between 0.3 and 0.5 of the
whole surface of the star. As the adopted configuration of spots is
not unique we do not think it worthwhile to go into details, and we
limit ourselves to just one example illustrated in
Fig.~\ref{fig:modul} which shows periodic variations in the $V$-band
caused solely by the presence of spots (\ie without the effect of
ellipticity). The data for Fig.~\ref{fig:modul} are taken from
models for which the base fit was obtained from the LCO photometry.
\begin{figure}[h!]
 \begin{center}
  \includegraphics[width=\textwidth, bb = 36 358 565 690, clip] {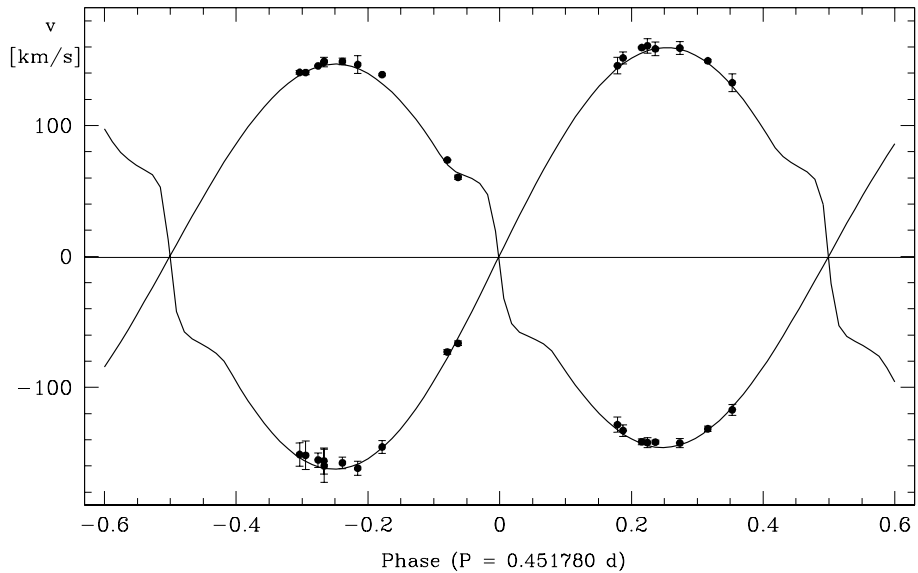}
  \caption {Example fit to UVES velocity data. The horizontal line shows
   the systemic velocity of $-0.8\pm0.4$ km s$^{-1}$.}
  \label{fig:velfit}
 \end{center}
 \vskip -5mm
\end{figure}

An example velocity fit for phases $\phi<0.9$ is shown in Fig.
\ref{fig:velfit} (bends on the synthetic curve are due to the
Rossiter - McLaughlin effect). We decided to not include points with
$\phi>0.95$ despite their formal errors in Table \ref{tab:radvel}
being rather small because we suspected that heavy line-blending in
the corresponding spectra would introduce unaccountable systematic
deviations. Indeed, after a few trial fits with all velocity points
taken into account we found that for $\phi>0.95$ it was impossible
to get residual velocities smaller than $10$ km s$^{-1}$, whereas
the rms residual velocity for $\phi<0.9$ was equal to only 2.1 km
s$^{-1}$.
\vskip -2 mm
\begin{table}[h!]
 \begin{center}
 \caption{Physical parameters of S892 \label{tab:syspar}}
  {\small
   \begin{tabular}{llrlr}
    \hline
    $P$ &[d]                   &0.451780&$\pm$0.000001&\\
    $a$ &[$R_\odot$]           &2.77    &$\pm$0.01    &$(^{+0.013}_{-0.015})$\\
    $e$ &(fixed)               &0.      &             &\\
    $i$ &[degrees]             &85.36   &$\pm$0.28    &$(^{+0.58}_{-0.24})$\\
    $\gamma$ &[km s$^{-1}$]    &-0.8    &$\pm$0.4     &\\
    $m_{\rm 1}$ &[$M_\odot$]   &0.733   &$\pm$0.005   &$(^{+0.011}_{-0.007})$\\
    $m_{\rm 2}$ &[$M_\odot$]   &0.662   &$\pm$0.005   &$(^{+0.012}_{-0.008})$\\
    $R_{\rm 1}$ &[$R_\odot$]   &0.719   &$\pm$0.014   &$(^{+0.030}_{-0.017})$\\
    $R_{\rm 2}$ &[$R_\odot$]   &0.680   &$\pm$0.017   &$(^{+0.031}_{-0.014})$\\
    \hline
    $T_{\rm 1}$ &[K] (fixed)   &4200    &        &\\
    $T_{\rm 2}$ &[K]           &3940    &$\pm$20      &$(^{+110}_{-100})$ \\
    ${M_{\rm bol}}_1$ &[mag]   &6.89    &$\pm$0.04    &$(^{-0.20}_{+0.15})$ \\
    ${M_{\rm bol}}_2$ &[mag]   &7.29    &$\pm$0.04    &$(^{-0.20}_{+0.16})$ \\
    \hline
   \end{tabular}
  }
 \end{center}
 \vskip -1 mm
 {\footnotesize 3$^{\rm rd}$ column: rms deviations from mean values
 obtained in 9 fits to light and velocity curves. 4$^{\rm th}$
 column above the horizontal line: largest deviations from the mean;
 below the horizontal line: variations corresponding to $\pm100$ K
 change in $T_{\rm 1}$
 }
\end{table}
\noindent All fits made according to the procedure described above
produced 9 sets of system parameters whose averaged values  are
listed in Table \ref{tab:syspar} together with rms deviations from
the nine-fit mean (since the "de-spotting" procedure may in
principle introduce systematic errors, we also give the largest
deviations from the mean). Obviously, both the mean values and the
errors might change if we had more light curves. Unfortunately, a
significant improvement of their accuracy would require at least
$\sim$10 additional light curves at epochs separated by at least
several weeks. Note that temperatures -- and, obviously, the
bolometric magnitudes they yield -- are not well constrained by the
data (see Section \ref{sect:dis}). Their errors in column 3 of Table
\ref{tab:syspar} are but formal results of the fits with $T_{\rm
1}=4200$ K; depending on the exact value of $T_{\rm 1}$ they may
vary as indicated in the last column.

To check the influence of photometry errors on values given in Table
\ref{tab:syspar} we removed spot effects from our best (LCO) and
poorest (CTIO) data, and performed extensive (10000 points) Monte
Carlo fitting with the help of the JKTEBOP code (Southworth \etal
2007). For the LCO light curve the errors of $R_1$ and $R_2$ were
equal to 0.0021 $R_\odot$ and 0.0026 $R_\odot$, respectively, while
for the CTIO light curve they amounted to 0.0075 $R_\odot$ and
0.0090 $R_\odot$. Since photometry errors $\sigma_{\rm ph}$ and
``de-spotting'' errors $\sigma_{\rm ds}$ are independent, the total
error may be approximated by the square root of the sum
($\sigma_{\rm ph}^2 + \sigma_{\rm ds}^2$). Such an operation causes
the errors of $R_{\rm 1}$ and $R_{\rm 2}$ to increase, respectively,
by 14\% and 13\% compared to the values given in the 3$^{\rm rd}$
column of Table \ref{tab:syspar}.

The radii of the components imply rotational velocities $v_{\rm
1}\sin i=80.8\pm1.1$~km~s$^{-1}$ and $v_{\rm 2}\sin
i=76.3\pm2.3$~km~s$^{-1}$, which agree very well with $v_{\rm 1}\sin
i=79.2\pm5.9$~km~s$^{-1}$ and $v_{\rm 2}\sin
i=76.2\pm8.8$~km~s$^{-1}$ derived from BF fits in Section
\ref{sect:spectr}, thus indicating the consistency of our solution.
Note also that the low systemic velocity of S892 adds credence to
our assumption concerning its metallicity.
%
%
%
%
\vskip -3 mm
\section {Discussion} \label{sect:dis}
\begin{figure}[t]
 \begin{center}
  \includegraphics[width=\textwidth, bb = 36 378 565 690, clip] {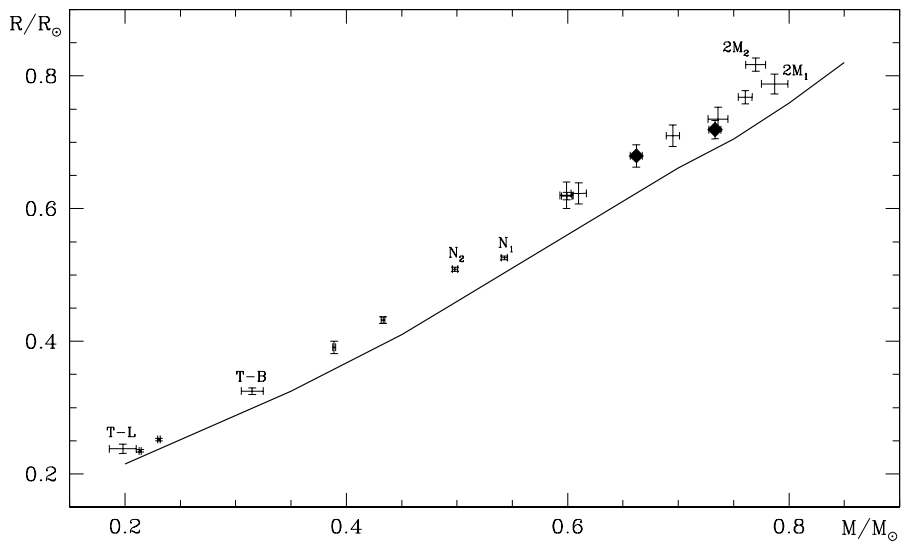}
  \caption
  {Mass-Radius diagram for low-mass eclipsing binary stars with observationally
   determined parameters (only stars with uncertainties below 3\% both in mass
   and radius are included). The components of S892 are indicated by diamonds.
   The solid line is a theoretical 1.0 Gyr isochrone based on the solar
   composition models of Baraffe et al. (1998). The data for NSVS 01031772
   (marked with N$_{1,2}$), 2MASS J05162881+2607387 (marked with 2M$_{1,2}$),
   T-Boo0-00080 (marked with T-B) and T-Lyr1-01662 (marked with T-L) are taken,
   respectively, from  L\'opez-Morales et al. (2006), Bayless \& Orosz (2006)
   and Fernandez \etal (2009). All remaining data (from left to right for
   CM~Dra~B, CM~Dra~A, CU~Cnc~B, CU~Cnc~A, GU~Boo~B, YY~Gem~AB, GU~Boo~A,
   RXJ0239.1-1028~B, RXJ0239.1-1028~A and V818~Tau~B) are taken from Ribas (2006).
  }
  \label{fig:lmms}
 \end{center}
 \vspace{-5mm}
\end{figure}

Several authors, \eg Morales \etal (2008) and references therein,
have found that radii of components of lower main sequence binary
stars (LMSS) are systematically and significantly larger than
predicted by the theory. Our analysis of S892 confirms this finding.
Incidentally, the radii of the components of this system agree very
well with the empirical $R(M)$ relation
\begin{equation}
 R/R_\odot = 0.0324 + 0.9343M/M_\odot
                 + 0.0374\left(M/M_\odot\right)^2
 \label{eq:bo}
\end{equation}
obtained by Bayless \& Orosz (2006), which for $M_{\rm 1}=0.733
M_\odot$ and $M_{\rm_2 }=0.662 M_\odot$ yields $R_{\rm
1}=0.737R_\odot$ and $R_{\rm_2 }=0.667R_\odot$. The
updated $M$--$R$ diagram for components of well-surveyed binary LMSS
is shown in Fig. \ref{fig:lmms}.

\begin{figure}[t!]
 \begin{center}
  \includegraphics[width=\textwidth, bb = 36 381 565 690, clip] {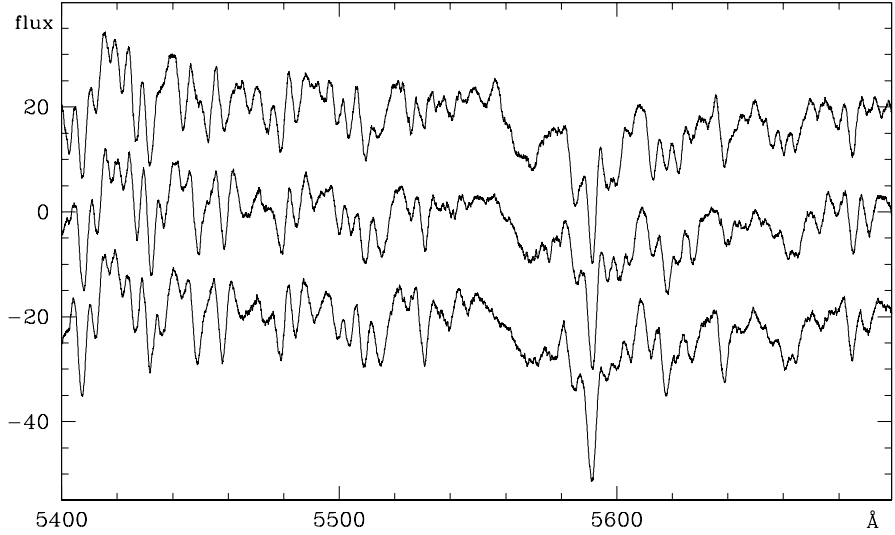}
  \caption {An illustration of the spectral variability of S892.
   From top to bottom: spectra obtained on HJD 2454746 ($\phi=0.177$), HJD 2454849
   ($\phi=0.695$) and HJD 2454744 ($\phi=0.723$). The flux scale is arbitrary; the
   top and bottom spectra are displaced vertically by 20 units.}
  \label{fig:spectvar}
 \end{center}
\end{figure}

As we already mentioned, based on our data it is impossible to find
accurate values of $T_{\rm 1}$ and $T_{\rm 2}$. This is because we
cannot derive absolute fluxes or effective temperatures from light
and velocity data alone as we do not know the distance to S892 from
independent measurements. In principle, we might try to match the
spectra of this system to binary spectral templates following Becker
et al. (2008). However, an additional problem arises here due to
large and variable spot coverage of both components. As a consequence,
the spectra of the components
are intrinsically variable (see Fig. \ref{fig:spectvar}), which puts
any matching attempts in question. Some changes in Fig.
\ref{fig:spectvar} are due to orbital motion; however most of them
(\eg at $5460<\lambda$[\AA]$<5480$, $5530<\lambda$[\AA]$<5590$ and
$\lambda$[\AA]$>5600$) result from the intrinsic variability of the components.
\begin{figure}[h!]
 \begin{center}
  \vspace{-2mm}
  \includegraphics[width=\textwidth, bb = 36 358 565 690, clip] {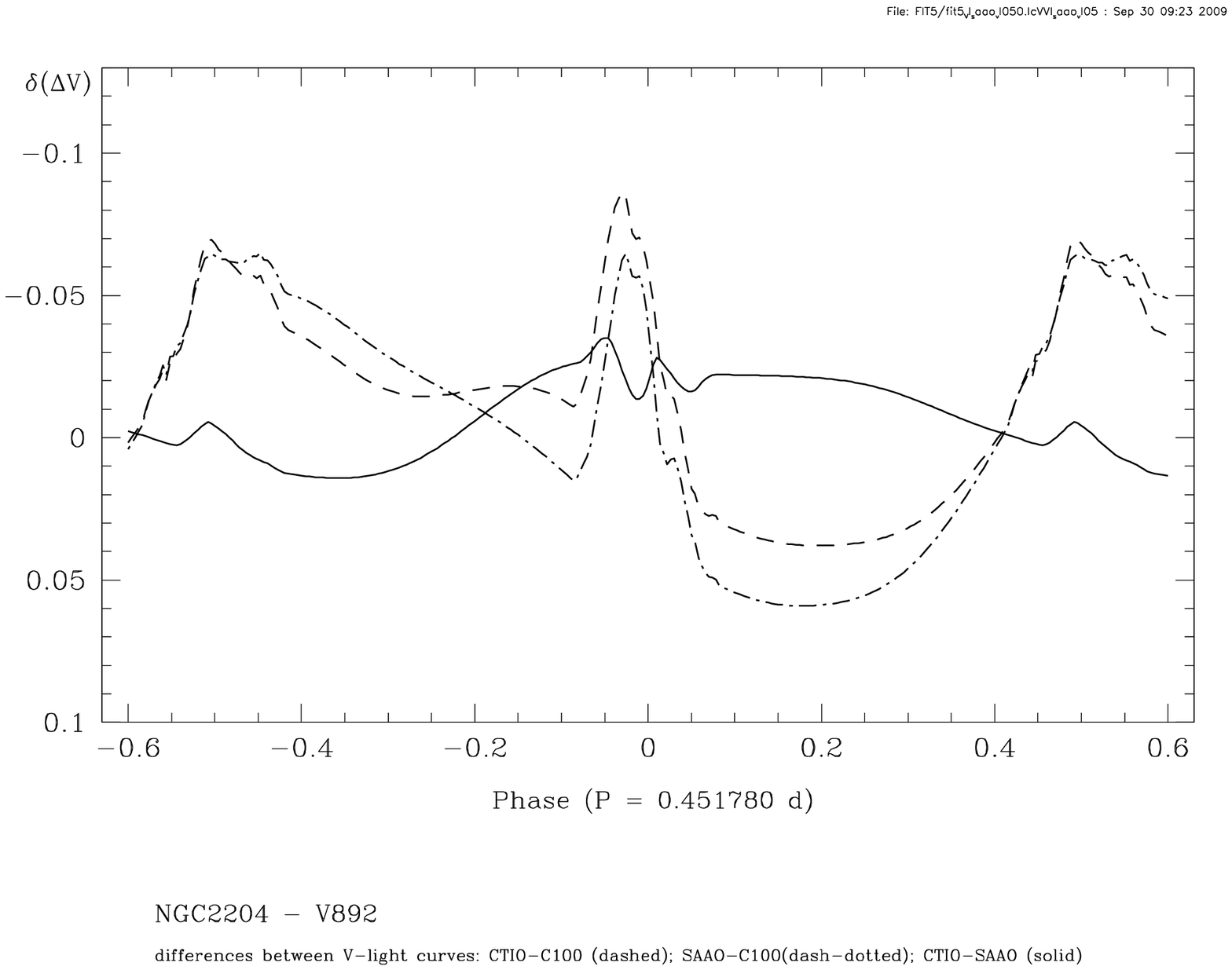}
  \caption {Variability of the $V$-band light curve illustrated by differences
   between solid curves from Fig. \ref{fig:Vfit}. Dashed: CTIO-LCO; dash-dotted:
   SAAO-LCO; solid: CTIO-SAAO. }
  \label{fig:Vdiffs}
  \vspace{-6mm}
 \end{center}
\end{figure}

\begin{figure}[t!]
 \begin{center}
  \includegraphics[width=\textwidth, bb = 54 432 565 690, clip] {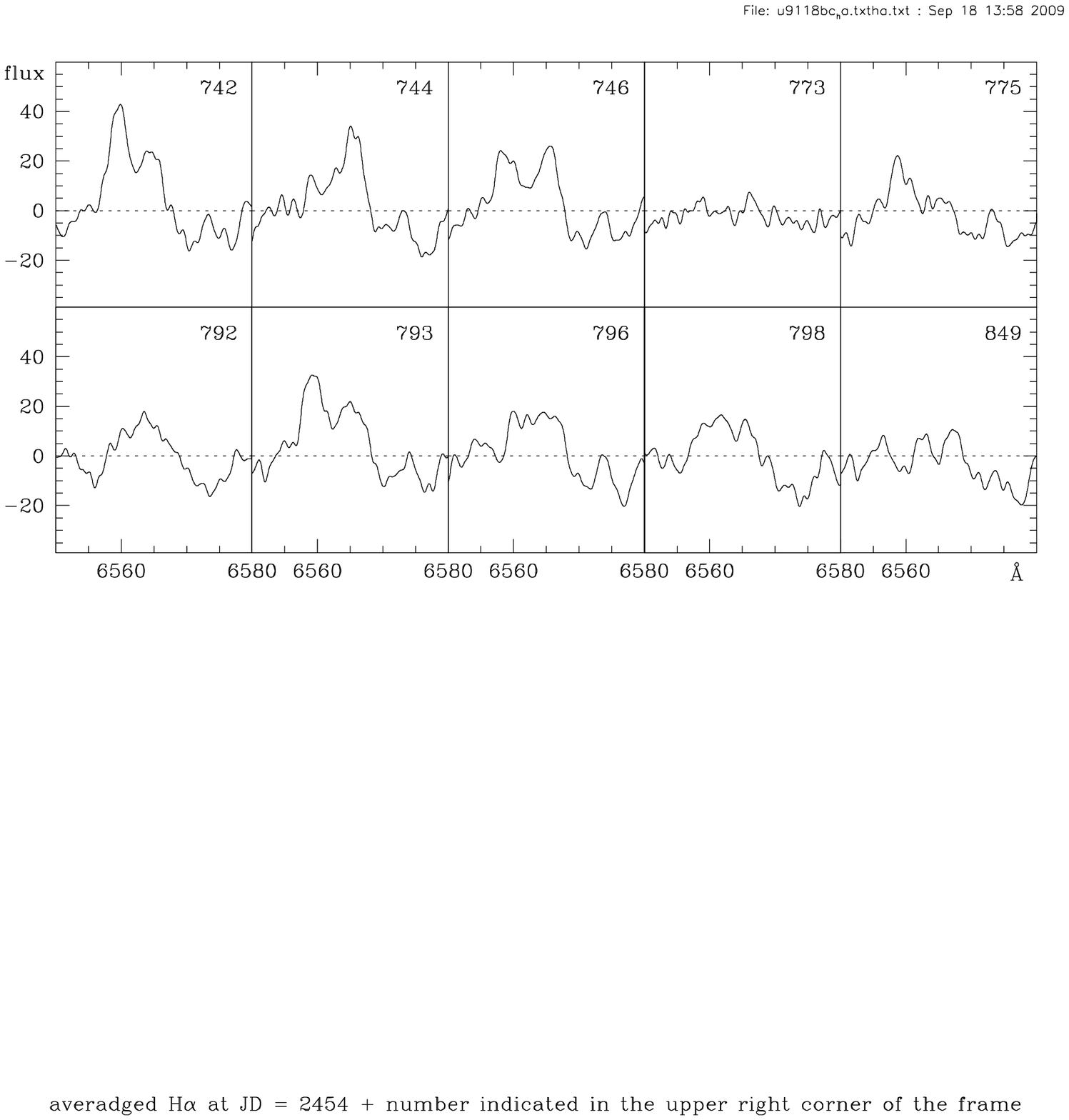}
  \caption {H$\alpha$ emission in S892. The equivalent width of the line
   varies from 0 to $\sim7${\AA} per component. The number in the upper
   right corner of each frame indicates the date of observation, in HJD-2454000.}
  \label{fig:halpha}
 \end{center}
 \vspace{-4mm}
\end{figure}

Another consequence of the vigorous activity of S892 is strong
variability of the light curve, as illustrated in Fig.
\ref{fig:Vdiffs}. One can see that at some orbital phases the
luminosity of the system may vary by almost 10\% within a year (note
that apparent phase shifts in Fig. \ref{fig:Vdiffs} are solely due
to asymmetries of the light curves). Yet another aspect of the
activity is strongly variable H$\alpha$ emission from both
components, with the equivalent width of the line changing within a
few weeks from practically 0 to $\sim7${\AA} per component (Fig.
\ref{fig:halpha}). H$\beta$ is also seen in emission, with the
equivalent width reaching up to $\sim3.5${\AA}. Neither H$\alpha$
nor H$\beta$ are ever seen in absorption. We were not able to detect
any secular or periodic intensity changes of either line, and we
concluded that the hydrogen emission seems to be entirely chaotic.

A system like this should be a fairly strong source of X-rays. Based
on the relation between radius and $L_{\rm X}/L_{\rm bol}$ found by
L\'opez-Morales (2007) we can expect the total X-ray luminosity of
S892 to reach $\sim5\times10^{29}$~erg~s$^{-1}$. Unfortunately,
because of the large distance of the system from the Sun the
expected flux ($\sim5\times10^{-15}$~erg~cm$^{-2}$~s$^{-1}$) is far
too low to be registered in the ROSAT all-sky survey. We did not
find relevant X-ray data in other catalogs or databases, either. An
attempt to estimate the age of S892 based on the equivalent width of
the Li~{\small I} $\lambda$6708 line was also futile -- the line
could not be detected. The quality of our spectra is insufficient to
draw any firm conclusion from this negative result -- we can only
say that the system almost certainly is not very young. Finally, we
searched for proper motion data. S892 is listed in the UCAC3
Catalogue (Finch et al. 2009) as 3UC143-028414, with
$\mu_\alpha=-7.6$~mas~yr$^{-1}$ and $\mu_\delta=-3.8$~mas~yr$^{-1}$,
but large errors (6.4~mas~yr$^{-1}$ in both coordinates) make this
information rather uninteresting.

Based on composite CaH$_2$, CaH$_3$ and TiO$_5$ molecular indices,
Becker et al. (2008) derived $T_{\rm eff}=3730\pm100$~K for the $M =
0.66$ $M_\odot$ primary component of 2MASS J01542930+0053266. While
this result marginally agrees with our $T_{\rm 2}$,\footnote {Note
that the larger error margin from the 4$^{\rm th}$ column of Table
\ref{tab:syspar} should be applied here.} it is tempting to verify
it using an additional evidence. Unfortunately, the indices used by
Becker et al. (2008) are defined for $\lambda>6800${\AA} (Reid,
Hawley \& Gizis 1995), \ie for wavelengths longer than the red
limit of our spectra. Not being able to apply their method to our
data, we compared their result to temperatures of similar systems
reported in the literature.

To our surprise, both stars in YY Gem and both stars in GU Boo, all
being by $\sim$10\% less massive than the primary of 2MASS~
J01542930+0053266 were found to be by up to ~350 K {\em hotter}
(Torres \& Ribas 2002; L\'opez-Morales \& Ribas 2005). Active stars
in this mass range are known to be cooler than nonactive ones (see
Section \ref{sect:intro}), but  in all three systems the level of
activity is similar (in principle, in 2MASS~J01542930+0053266 it
should be even lower than in the remaining two because of a
relatively long orbital period). Adopting $T_{\rm 2}=3730$~K for our
secondary and using the flux ratio obtained from our fits, we obtain
for our 0.73 $M_\odot$ primary $T_{\rm 1}=3940$~K -- a value
assigned by L\'opez-Morales \& Ribas (2005) to GU Boo A, whose mass
is smaller by almost 20\%. We conclude that 3940~K from Table
\ref{tab:syspar} is a fairly realistic value of $T_{\rm 2}$ in the
S892 system. Note that while all these discrepancies provide a good
illustration of the notorious problem with temperature measurements,
they luckily do not influence our main findings.

Bolometric corrections $BC_{\rm 1} = -1.00$~mag and $BC_{\rm 2} =
-1.25$~mag (Bessell, Castelli \& Plez 1998) translate our ${M_{\rm
bol}}_{\rm 1}$ and ${M_{\rm bol}}_{\rm 2}$ into ${M_{\rm V}}_{\rm 1}
= 7.89$~mag and ${M_{\rm V}}_{\rm 2} = 8.54$~mag, resulting in a
combined absolute $V$-band magnitude of 7.41~mag at the phase of
maximum light. NGC2204 is ~$\sim4$ kpc distant (Kassis \etal 1997)
and only weakly reddened, with $E(B-V)\approx0.1$ (Schlegel \etal
1998) and $A_V\approx0.32$. Thus, the revised distance to S892 is
about $1050$~pc if the whole absorbing material is located behind
the system, and about $900$ pc if it is located in front of it.
\section {Conclusions}
We discovered and analyzed a detached eclipsing binary composed of
two late K dwarfs. Based on light and velocity curves we derived
orbital period of 0.451780$\pm$0.000001 d, semimajor axis
$a=2.77\pm0.01$ $R_\odot$ and inclination $i$=85\zdot$^\circ$36
$\pm$0\zdot$^\circ$28. The masses of the components ($M_1 = 0.733 \pm
0.005$ $M_\odot$; $M_2 = 0.662 \pm 0.005$ $M_\odot$) and their radii
($R_1 = 0.72 \pm 0.01$ $R_\odot$; $R_2 = 0.68 \pm 0.02$ $R_\odot$)
are consistent with the empirical mass-radius relationship
established recently for LMSS in binary systems by Bayless \& Orosz
(2006): both stars are larger and cooler compared to theoretical
models.

The discrepancy between theory and observations was originally
thought to arise from inaccuracies in the equation of state and/or
opacity. It turned out, however, that models based on the same
physics were entirely compatible with observed properties of single
stars or stars in wide binaries (L\'opez-Morales et al. 2006).
Another approach to the problem was suggested by a positive
correlation between radius and activity level found for a few
members of close binaries by L\'opez-Morales (2007). It was realized
that the basic difference between single stars and components of
short-period binary systems was high rotational velocities of the
latter, promoting high levels of magnetic activity. Strong magnetic
fields can inflate the star due to large spot coverage (thus
lowering the effective temperature) and/or partial inhibition of
convection (which has the same effect as increased opacity; see
Chabrier \etal for details).

Both components of our binary are very active, showing high
spot-coverage, unstable light curves, and strong variability of the
emission in hydrogen lines. Thus, our results support the increasing
evidence that the observed inflation of radii of K and M dwarfs is
related to high levels of magnetic activity.
\Acknow{JK, PP, WP and MR were supported by the grant MISTRZ from
the Foundation for the Polish Science and by the grant N N203 379936
from the Polish Ministry of Science and Higher Education (PMSHE). MC
and PP acknowledge support from Chilean FONDAP Centro de
Astrof\'{i}sica No. 15010003. Support for PP was also provided by
the grant N N203 301355 from PMSHE, and MC was also supported by
Proyecto Basal PFB-06/2007 and Proyecto FONDECYT Regular \#1071002.
IBT acknowledges support from NSF grant\#0507325. This research made
use of the databases SIMBAD (operated at CDS, Strasbourg, France)
and WEBDA (operated at the Institute for Astronomy of the University
of Vienna). We thank the staff of VLT for the crucial spectroscopic
observations, and the referee, Grzegorz Pojmanski, for his
constructive remarks.}

\begin{references}
 \refitem{Baraffe I., Chabrier G., Allard F. and  Hauschildt P. H.}
        {1998}{\AA}{337}{403}
 \refitem{Bessell M.S., Castelli F. and Plez B.}{1998}{\AA}{333}{231}
 \refitem{Bayless A.J. and Orosz J.A.}{2006}{\ApJ}{651}{1155}
 \refitem{Becker A.C., Agol E., Silvestri N.M., Bochanski J.J., Laws
        C., West A.A., Basri G., Belokurov V., Bramich D.M., Carpenter J.M.,
        Challis P., Covey K.R., Cutri R.M., Evans N.W., Fellhauer M., Garg
        A., Gilmore G., Hewett P., Plavchan P., Schneider D.P., Slesnick
        C.L., S. Vidrih S., Walkowicz L.M. and Zucker D.B.}{2008}
        {\MNRAS}{386}{416}
 \refitem{Blake C.H., Torres G., Bloom J.S. and Gaudi B.S.}
        {2008}{\ApJ}{684}{635}
 \refitem{Chabrier, G., Gallardo, J. and Baraffe, I.}{2007}{\AA}{472}{L17}
 \refitem{Claret, A.}{2000}{\AA}{359}{289}
 \refitem{Demory B.-O, S\'egransan D., Forveille T., Queloz D.,
        Beuzit J.-L., Delfosse X., Di Folco E., Kervella P.,
        Le Bouquin J.-B., and Perrier C.}{2009}{\AA}{505}{205}
 \refitem {Fernandez J.M., Latham D.W., Torres G., Everett M.E., Mandushev G.,
        Charbonneau D., O¿Donovan F.T., Alonso R., Esquerdo G.A., Hergenrother C.W.
        and Stefanik R.P.}{2009}{\ApJ}{701}{764}
 \refitem {Finch C.T., Zacharias N., Girard T., Wycoff G. and Zacharias M.I.}
        {2009}{Bull. Am. Ast. Soc.}{41}{427}
 \refitem{Kassis M., Janes K.A., Friel E.D. and Phelps R.L.}
        {1997}{\AJ}{113}{1723}
 \refitem{L\'opez-Morales M.}{2007}{\ApJ}{660}{732}
 \refitem{L\'opez-Morales M., Orosz J.A., Shaw J.S., Havelka L.,
         Ar\'evalo M.J., McIntyre T. and L\'azaro C.}{2006}
         {astro-ph/06110225}{~}{~}
 \refitem{L\'opez-Morales M. and Ribas I.}{2005}{\ApJ}{631}{1120}
 \refitem{Lucy, L.}{1967}{Zeitschrift für Astrophysik}{6}{89}
 \refitem{Morales J.C., Ribas I. and Jordi C.}{2008}{\AA}{478}{507}
 \refitem{Mohanty S. and Basri G.}{2003}{\ApJ}{583}{451}
 \refitem{Pr\v sa A. and Zwitter T.}{2005}{\AJ}{628}{426}
 \refitem{Reid I.N., Hawley S.L. and Gizis J.E.}{1995}{\AJ}{110}{1838}
 \refitem{Ribas I.}{2006}{Astrophys.Sp.Sci.}{304}{89}
 \refitem{Ribas I., Morales J.C., Jordi C., Baraffe I., Chabrier
        G. and Gallardo J.}{2008}{Mem.Soc.Astr.It.}{79}{562}
 \refitem{Rozyczka M., Kaluzny J., Krzeminski W. and Mazur B.}{2007}
        {\Acta}{57}{323}
 \refitem{Rucinski S.M.}{2002}{\AJ}{124}{1746}
 \refitem{Schlegel, D.J., Finkbeiner, D.P. and Davis,
        M.}{1998}{\ApJ}{500}{525}
 \refitem{Shaw J.S. and L\'opez-Morales M.}{2007}{ASPC}{362}{15}
 \refitem{Southworth, J., Bruntt, H. and Buzasi, D. L.}{2007}{\AA}{467}{1215}
 \refitem{Stassun K.G., Hebb L., L\'opez-Morales M. and Pr\v sa A.}
        {2009}{IAU Symp}{258}{161}
 \refitem{Stetson P.B.}{1987}{\PASP}{99}{191}
 \refitem{Torres G. and Ribas I.}{2002}{\ApJ}{567}{1140}
 \refitem{Wilson R.E. and Devinney E.J.}{1971}{\AJ}{166}{605}
\end{references}
\end{document}